\RequirePackage{fixltx2e}
\documentclass[nofootinbib,reprint,showpacs,noeprint,
  prc,aps,10pt,superscriptaddress,floatfix]{revtex4-1}
\usepackage{graphicx}
\usepackage{hyperref}
\usepackage{epstopdf}
\usepackage{amsmath}
\usepackage{amssymb}
\usepackage[amsmath,thmmarks]{ntheorem}
\usepackage{bm}
\usepackage{cleveref}
\usepackage{subfigure}
\usepackage{epstopdf}
\usepackage{array}
\usepackage{ifthen}
\usepackage{booktabs}
\usepackage{textgreek}
\usepackage{multirow}
\usepackage{enumitem}
\usepackage{dcolumn}
\usepackage{comment}
\newcolumntype{C}[1]{>{\centering\let\newline\\\arraybackslash\hspace{0pt}}m{#1}}

\let\oldtabular\tabular
\renewcommand{\tabular}{\large\oldtabular}

\usepackage{color}
 
\graphicspath{{}}

\begin{document}

\title{Search for Majoron-emitting modes of double-beta decay of {$^{136}$Xe} with EXO-200}

\newcommand{\IHEP}{\affiliation{Institute of High Energy Physics, Beijing, China}}
\newcommand{\DukeTUNL}{\affiliation{Department of Physics, Duke University, and Triangle Universities Nuclear Laboratory (TUNL), Durham, North Carolina 27708, USA}}
\newcommand{\TRIUMF}{\affiliation{TRIUMF, Vancouver, British Columbia V6T 2A3, Canada}}
\newcommand{\Seoul}{\affiliation{Department of Physics, University of Seoul, Seoul, Korea}}
\newcommand{\Stanford}{\affiliation{Physics Department, Stanford University, Stanford, California 94305, USA}}
\newcommand{\WIPP}{\affiliation{Waste Isolation Pilot Plant, Carlsbad, New Mexico 88220, USA}}
\newcommand{\Laurentian}{\affiliation{Department of Physics, Laurentian University, Sudbury, Ontario P3E 2C6, Canada}}
\newcommand{\ITEP}{\affiliation{Institute for Theoretical and Experimental Physics, Moscow, Russia}}
\newcommand{\Illinois}{\affiliation{Physics Department, University of Illinois, Urbana-Champaign, Illinois 61801, USA}}
\newcommand{\Caltech}{\affiliation{Kellogg Lab, Caltech, Pasadena, California 91125, USA}}
\newcommand{\UMass}{\affiliation{Physics Department, University of Massachusetts, Amherst, Massachusetts 01003, USA}}
\newcommand{\TUM}{\affiliation{Technische Universit\"at M\"unchen, Physikdepartment and Excellence Cluster Universe, Garching, Germany}}
\newcommand{\Maryland}{\affiliation{Physics Department, University of Maryland, College Park, Maryland 20742, USA}}
\newcommand{\LHEP}{\affiliation{LHEP, Albert Einstein Center, University of Bern, Bern, Switzerland}}
\newcommand{\CSU}{\affiliation{Physics Department, Colorado State University, Fort Collins, Colorado 80523, USA}}
\newcommand{\Indiana}{\affiliation{Physics Department and CEEM, Indiana University, Bloomington, Indiana 47405, USA}}
\newcommand{\Alabama}{\affiliation{Department of Physics and Astronomy, University of Alabama, Tuscaloosa, Alabama 35487, USA}}
\newcommand{\Drexel}{\affiliation{Department of Physics, Drexel University, Philadelphia, Pennsylvania 19104, USA}}
\newcommand{\SLAC}{\affiliation{SLAC National Accelerator Laboratory, Stanford, California 94025, USA}}
\newcommand{\Carleton}{\affiliation{Physics Department, Carleton University, Ottawa, Ontario K1S 5B6, Canada}}
\author{J.B.~Albert}\Indiana
\author{D.J.~Auty}\Alabama
\author{P.S.~Barbeau}\DukeTUNL
\author{E.~Beauchamp}\Laurentian
\author{D.~Beck}\Illinois
\author{V.~Belov}\ITEP
\author{C.~Benitez-Medina}\altaffiliation{Now at Intel, Hillsboro, OR, USA}\CSU
\author{M.~Breidenbach}\SLAC
\author{T.~Brunner}\Stanford
\author{A.~Burenkov}\ITEP
\author{G.F.~Cao}\IHEP
\author{C.~Chambers}\CSU
\author{J.~Chaves}\Stanford
\author{B.~Cleveland}\altaffiliation{Also SNOLAB, Sudbury ON, Canada}\Laurentian
\author{M.~Coon}\Illinois
\author{A.~Craycraft}\CSU
\author{T.~Daniels}\UMass
\author{M.~Danilov}\ITEP
\author{S.J.~Daugherty}\Indiana
\author{C.G.~Davis}\altaffiliation{Now at the Naval Research Lab, Washington D.C., USA}\Maryland
\author{J.~Davis}\Stanford
\author{R.~DeVoe}\Stanford
\author{S.~Delaquis}\LHEP
\author{T.~Didberidze}\Alabama
\author{A.~Dolgolenko}\ITEP
\author{M.J.~Dolinski}\Drexel
\author{M.~Dunford}\Carleton
\author{W.~Fairbank Jr.}\CSU
\author{J.~Farine}\Laurentian
\author{W.~Feldmeier}\TUM
\author{P.~Fierlinger}\TUM
\author{D.~Fudenberg}\Stanford
\author{G.~Giroux}\altaffiliation{Now at Queen's University, Kingston, ON, Canada}\LHEP
\author{R.~Gornea}\LHEP
\author{K.~Graham}\Carleton
\author{G.~Gratta}\Stanford
\author{C.~Hall}\Maryland
\author{S.~Herrin}\SLAC
\author{M.~Hughes}\Alabama
\author{M.J.~Jewell}\Stanford
\author{X.S.~Jiang}\IHEP
\author{A.~Johnson}\SLAC
\author{T.N.~Johnson}\Indiana
\author{S.~Johnston}\UMass
\author{A.~Karelin}\ITEP
\author{L.J.~Kaufman}\Indiana
\author{R.~Killick}\Carleton
\author{T.~Koffas}\Carleton
\author{S.~Kravitz}\Stanford
\author{A.~Kuchenkov}\ITEP
\author{K.S.~Kumar}\UMass
\author{D.S.~Leonard}\Seoul
\author{F.~Leonard}\Carleton
\author{C.~Licciardi}\Carleton
\author{Y.H.~Lin}\Drexel
\author{J.~Ling}\Illinois
\author{R.~MacLellan}\SLAC
\author{M.G.~Marino}\TUM
\author{B.~Mong}\Laurentian
\author{D.~Moore}\Stanford
\author{R.~Nelson}\WIPP
\author{A.~Odian}\SLAC
\author{I.~Ostrovskiy}\email[Corresponding author: ]{ostrov@stanford.edu}\Stanford
\author{C.~Ouellet}\Carleton
\author{A.~Piepke}\Alabama
\author{A.~Pocar}\UMass
\author{C.Y.~Prescott}\SLAC
\author{A.~Rivas}\Stanford
\author{P.C.~Rowson}\SLAC
\author{M.P.~Rozo}\Carleton
\author{J.J.~Russell}\SLAC
\author{A.~Schubert}\Stanford
\author{D.~Sinclair}\Carleton\TRIUMF
\author{E.~Smith}\Drexel
\author{V.~Stekhanov}\ITEP
\author{M.~Tarka}\Illinois
\author{T.~Tolba}\LHEP
\author{D.~Tosi}\altaffiliation{Now at University of Wisconsin, Madison, WI, USA}\Stanford
\author{R.~Tsang}\Alabama
\author{K.~Twelker}\Stanford
\author{P.~Vogel}\Caltech
\author{J.-L.~Vuilleumier}\LHEP
\author{A.~Waite}\SLAC
\author{J.~Walton}\Illinois
\author{T.~Walton}\CSU
\author{M.~Weber}\Stanford
\author{L.J.~Wen}\IHEP
\author{U.~Wichoski}\Laurentian
\author{L.~Yang}\Illinois
\author{Y.-R.~Yen}\Drexel
\author{O.Ya.~Zeldovich}\ITEP

\collaboration{EXO-200 Collaboration}
\noaffiliation

\date{\today}

\begin{abstract}

EXO-200 is a single phase liquid xenon detector designed to search for neutrinoless 
double-beta decay of \(^{136}\)Xe. 
Here we report on a search for various Majoron-emitting modes based on 100 kg\(\cdot\)yr exposure of  \(^{136}\)Xe. 
A lower limit of \(T^{^{136}Xe}_{1/2}>\) 1.2\(\cdot\)10\(^{24}\) yr at 90\% C.L. on the half-life of the spectral index = 1 Majoron decay was obtained, corresponding to a constraint on the Majoron-neutrino coupling constant of \(|\langle g^{M}_{ee} \rangle| <\) (0.8-1.7)\(\cdot\)10\(^{-5}\).

\end{abstract}

\pacs{23.40.-s, 21.10.Tg, 14.60.Pq, 14.80.Va}

\maketitle

\section{Introduction}\label{sec:Introduction} 

Double-beta decay ($\beta\beta$) is a rare radioactive transition between two nuclei with the same mass number A and with the nuclear charges Z different by two units. The process
can only proceed when the initial even-even nucleus is less bound than the final one, and can only be observed when both are more bound than the intermediate odd-odd nucleus
(or the decay to the intermediate nucleus is highly suppressed, as in $^{48}$Ca). Thus, in $\beta\beta$ decay, two neutrons are transformed into two protons and two electrons
simultaneously, with or without the emission of additional neutral particles.

Several modes of the $\beta\beta$ decay are considered in the literature. 
The mode where two antineutrinos are emitted together with the electrons (the two neutrino decay $2\nu\beta\beta$) is an allowed
decay in the Standard Model that conserves total lepton number. This mode has been observed in several cases, in particular recently in $^{136}$Xe~\cite{Ackerman:2011gz, KamLANDZen:2012aa} with a half-life of $T^{2\nu\beta\beta} = 2.165 \pm 0.016(stat) \pm 0.059(sys) \times 10^{21}$ yr~\cite{Albert:2014}.  In contrast, there are alternative, so-far-unobserved, {\it{neutrinoless}} modes where
the total lepton number is not conserved and whose existence requires that neutrinos are massive Majorana particles~\cite{Schechter:1982}. The simplest
of such modes, the $0\nu\beta\beta$ decay with the emission of two electrons, and nothing else, is a subject of an intense experimental search. In particular, for $^{136}$Xe half-life limits have most recently been set to $>1.1 \cdot 10^{25}$ yr~\cite{Albert:2014b} and $>1.9 \cdot 10^{25}$yr~\cite{Gando:2012zm}.

In this work we derive
half-life limits for the neutrinoless modes of $^{136}$Xe in which one or two additional bosons, denoted as $\chi_0$ here, are emitted together with the electrons, schematically
\begin{align}
(A,Z) \rightarrow (A, Z+2) + 2e^- + \chi_0,
\end{align}
or
\begin{align}
(A,Z) \rightarrow (A, Z+2) + 2e^- + 2\chi_0.
\end{align}

The boson(s) emitted in the $0\nu\beta\beta\chi_0$ or $0\nu\beta\beta\chi_0\chi_0$ modes is (are) usually referred to as ``Majoron(s)''.
Originally described as a Goldstone boson associated with spontaneous lepton number symmetry breaking, Majorons are possible dark matter candidates~\cite{Lattanzi:2013} and may be involved in other cosmological and astrophysical processes (e.g.~\cite{Dolgov:2004,Das:2011}).
Although the original proposals by Gelmini and Roncadelli~\cite{Gelmini:1981} and Georgi et al.~\cite{Georgi:1981} are disfavored by precise measurement of the width of the Z boson decay to invisible channels~\cite{lep:2006}, other analogous models were proposed, free of this constraint, in which Majoron more generally
refers to massless or light bosons that might be neither Goldstone bosons, nor required to carry a lepton charge (see~\cite{Barabash:2004} and references therein). 

\begin{figure}[htbp]
 \centering
 \includegraphics[width=0.5\textwidth]{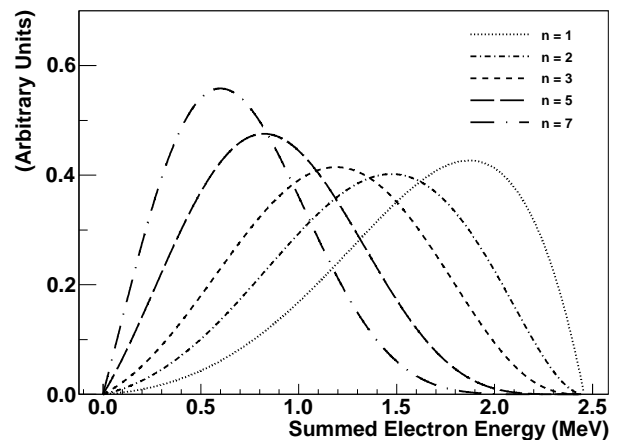}
  \caption{Spectra for the n=1,2,3, and 7 Majorons, as well as for the \(2\nu\beta\beta\) (n=5) decays of $^{136}$Xe.}
 \label{fig:spectra}
\end{figure}

The Majoron-emitting modes are experimentally recognizable by the shape of the sum electron spectrum \(S(E_{sum})\), characterized by the spectral index $n$,
\begin{widetext}
\begin{align}
\label{eq:shapes}
S(E_{sum}) = \int_{1}^{E_{sum}-1}F(Z,E_1)E_1p_1F(Z,E_2)E_2p_2(E_{tot}-E_1-E_2)^ndE_1dE_2\delta(E_{sum}-E_1-E_2) ~,
\end{align}
\end{widetext}
where $E_1$, $p_1$, $E_2$, and $p_2$ are the energy and momentum for each of the two electrons and $E_{sum} = E_1 + E_2$ is the observable sum energy, $E_{tot}$ is the total available energy, i.e. the decay Q value plus two electron masses, and the spectral index is an integer $n$ = 1, 2, 3, or 7. $F(Z,E)$ is the Fermi function that represents the effect of the nuclear (and atomic) Coulomb field on the wave function of the outgoing electron. All energies are in units of the electron mass $m_e$ and thus the function $S(E_{sum})$ is dimensionless. Note that $n$=5 for the observed $2\nu\beta\beta$ decay.
The normalized spectra for $^{136}$Xe and various spectral indices are illustrated in Figure~\ref{fig:spectra}.

It is beyond the scope of this paper to discuss the characteristic features of the different Majoron models that are discussed in ~\cite{Bamert:1995,Hirsh:1996}.
Generally, the half-life, effective Majoron-neutrino coupling constant $g_{\alpha}$, phase space integral, and the nuclear matrix elements $M^{\alpha}$ are related by
\begin{equation}\label{eq:coupling}
\frac{1}{T_{1/2}} = | \langle g_{\alpha} \rangle |^m \cdot |M'^{\alpha}|^2 \cdot G^{0\nu M}_{\alpha}(Z,E_0) ~,
\end{equation}
where \(M'^{\alpha} = M^{\alpha}\left(\frac{g_A}{1.25}\right)^2\), $g_A$ is the axial coupling constant, $m = 2(4)$ for the emission of one (two) Majorons, and $G^{0\nu M}_{\alpha}(Z,E_0)$ is the unnormalized phase space integral that depends on the  model type, $\alpha$ (see Table~\ref{tab:models}) and contains all the necessary fundamental constants. For completeness we show in Table~\ref{tab:models} the most important characteristics of ten Majoron models considered in recent experimental \(\beta\beta\) decay searches~\cite{Arnold:2006,KamLANDZen:2012zzg}. 

\begin{table*}[t]
\centering
\begin{tabular}{l@{\hskip 0.2in} c@{\hskip 0.2in} c@{\hskip 0.2in} c@{\hskip 0.2in} c@{\hskip 0.2in}}
\hline\hline
\parbox[t]{2.5cm} {Model type, \\\(\alpha\)} & \parbox[t]{2.5cm}{Number of Majorons emitted in \(0\nu\beta\beta\) decay, \(m\)} &\parbox[t]{3cm}{Is the Majoron a Goldstone boson?} & \parbox[t]{2cm}{Lepton charge,\\ \(L\)} &\parbox[t]{2cm}{Spectral index,\\ \(n\) }\\
\hline
IB	&	1 	&	no	&	0	&	1\\
IC	&	1	&	yes	&	0	&	1\\
ID	&	2	&	no	&	0	&	3\\
IE	&	2	&	yes	&	0	&	3\\
IIB	&	1	&	no	&	-2	&	1\\
IIC	&	1	&	yes	&	-2	&	3\\
IID	&	2	&	no	&	-1	&	3\\
IIE	&	2	&	no	&	-1	&	7\\
IIF	&	1	&	no	&	-2	&	3\\
``bulk''	&	1	&	no	&	0	&	2\\
\hline\hline
\end{tabular}
  \caption{Different Majoron-emitting models of \(0\nu\beta\beta\) decay. Class I (II) corresponds to lepton-number-violating (-conserving) models, with subclasses, denoted by letters, corresponding to different quantum numbers of a new particle (detailed description of the classification scheme in~\cite{Cowsik:1996tg, Bamert:1995, Hirsh:1996}).  
In the ``bulk'' model, built in the context of the brane-bulk scenarios for particle physics, the Majoron is a bulk singlet whose Kaluza-Klein excitations may make it visible in \(0\nu\beta\beta\) decay~\cite{Mohapatra:2000}.}
\label{tab:models}
\end{table*}

Table~\ref{tab:psfs} shows the phase space integrals for different values of the spectral index for $^{136}$Xe. 
For numerical calculations it is important to employ accurate values of the Fermi function $F(Z,E)$.
The $F(Z,E)$ used in preparing this table was calculated by a code~\cite{vogel_private:2014} that fully includes the nuclear finite size and electron screening and, as recently recommended~\cite{Kotila:2012zza}, evaluates F(Z,E) at the nuclear radius $R$.

\begin{table*}[t]
\centering
\begin{tabular}{c|ccccc}
\hline\hline
\parbox[t]{1.2cm}{Decay mode} &
\parbox[t]{2cm}{\(0\nu\beta\beta\chi_0\)\\ n=1 } &\parbox[t]{2cm}{\(0\nu\beta\beta\chi_0\) \\ n=3 } & \parbox[t]{2cm}{\(0\nu\beta\beta\chi_0\chi_0\) \\ n=3 } &\parbox[t]{2cm}{\(2\nu\beta\beta\) \\ n=5 } &\parbox[t]{2cm}{\(0\nu\beta\beta\chi_0\chi_0\) \\ n=7 }\\
\hline
& & & & & \\ 
\parbox[t]{1.2cm}{Const.}&
\large \(\frac{(G_F cos\theta_C g_A)^4 m^7_e (\hbar c)^2}{128\pi^7\hbar log(2) R^2}\) 
&\large  \(\frac{(G_F cos\theta_C g_A)^4 m^9_e}{32\pi^7\hbar log(2)}\) 
&\large  \(\frac{(G_F cos\theta_C g_A)^4 m^7_e (\hbar c)^2}{6114\pi^9\hbar log(2) R^2}\) 
&\large \(\frac{(G_F cos\theta_C g_A)^4 m^9_e}{240\pi^7\hbar log(2)}\) 
&\large  \(\frac{(G_F cos\theta_C g_A)^4 m^7_e (\hbar c)^2}{107520\pi^9\hbar log(2) R^2}\)\\
& & & & & \\
 \parbox[t]{1.2cm}{\(G^{0\nu M}_{\alpha}\)} &
\textbf{\parbox[t]{2cm}{1.11\(\cdot\)10\(^{-15}\)}}
&\textbf{\parbox[t]{2cm}{4.02\(\cdot\)10\(^{-18}\)}}
&\textbf{\parbox[t]{2cm}{8.32\(\cdot\)10\(^{-18}\)}}
&\textbf{\parbox[t]{2cm}{3.86\(\cdot\)10\(^{-18}\)}}
&\textbf{\parbox[t]{2cm}{3.44\(\cdot\)10\(^{-17}\)}}\\
\hline\hline
\end{tabular}
\caption{Phase space functions in yr$^{-1}$ for various Majoron
modes and for the $2\nu\beta\beta$ decay of $^{136}$Xe evaluated at nuclear radius $R = 1.2 A^{1/3}$ fm. The constants in front of the integral are also shown (where \(G_F\) is the Fermi constant and \(\theta_C\) is the Cabibbo angle).
The units are such that all energies in the integrals are in units of $m_e$.}
\label{tab:psfs}
\end{table*}

\section{Detector description}\label{sec:Detector} 

The EXO-200 detector is a cylindrical single phase time projection chamber (TPC) filled with liquid xenon enriched to 80.6\% in \(^{136}\)Xe. A detailed description of the detector is available elsewhere~\cite{Auger:2012gs}. The detector is constructed from components carefully selected to minimize internal radioactivity~\cite{Leonard:2007uv}. External radioactivity is shielded by 25 cm thick lead walls surrounding the detector on all sides. Additional passive shielding is provided by \(\sim\)50 cm of high purity cryogenic fluid~\cite{3m} filling the copper cryostat with a wall thickness of 5.4 cm that houses the TPC. The detector is located inside a clean room at the Waste Isolation Pilot Plant (WIPP) in Carlsbad, NM, USA, under an overburden of 1585\(^{+11}_{-6}\) meters water equivalent~\cite{Esch:2004zj}. The remaining cosmic ray flux is detected by an active muon veto system consisting of plastic scintillation panels surrounding the clean room on four sides. Energy deposited in the TPC by ionizing radiation produces free charge and scintillation light, which are registered by anode wire grids and arrays of avalanche photodiodes, respectively. The TPC allows for three-dimensional position reconstruction of energy depositions, providing further discrimination against gamma backgrounds. Charge deposits (clusters) in a given event that are spatially separated by \(\sim\)1 cm or more can be individually resolved. The event can then be classified as single-site (SS), or multi-site (MS), depending on the number of observed charge clusters. Based on Monte Carlo (MC) simulation, \textgreater90\% of \(\beta\beta\) events are expected to be reconstructed as SS, while the energy-averaged fraction of SS gamma events is around 30\%. Total energy of an event is determined by combining the charge and scintillation signals, which achieves better energy resolution than in each individual channel due to the anticorrelation between them~\cite{Conti:2003}. Radioactive gamma sources are periodically deployed at several positions near the TPC to characterize the detector response and validate the MC simulation.

\section{Experimental data and analysis}\label{sec:Analysis}

The data set and event selection criteria used in this work are the same as in the recent search for the neutrino mediated \(0\nu\beta\beta\) decay~\cite{Albert:2014b}. 
The data were collected between September 22, 2011 and September 1, 2013 resulting in the total of 477.60\(\pm\)0.01 live days. 
The fiducial volume is described by a hexagon with an apothem of 162 mm and absolute length coordinate values between 10 and 182 mm (with Z = 0 corresponding to the cathode location). This translates into a \(^{136}\)Xe mass of 76.5 kg, or 3.39\(\cdot\)10\(^{26}\) atoms of \(^{136}\)Xe, and an exposure of 100 kg\(\cdot\)yr (736 mol\(\cdot\)yr). 

\begin{figure}[htp]
 \centering
 \includegraphics[width=0.5\textwidth]{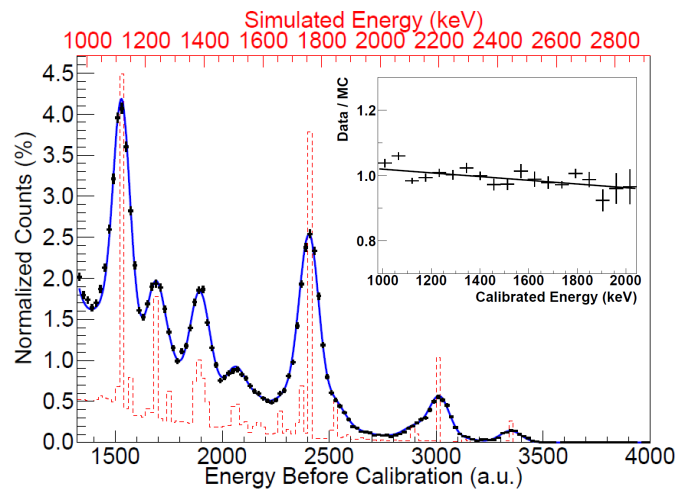}
  \caption{(Color online) Example of an energy spectrum fit using \(^{226}\)Ra data (black points) and corresponding MC simulation. The dotted line shows the MC energy spectrum before
the fit, without the detector effects of energy smearing and at the correct energy scale (indicated by the upper scale in red). The continuous line depicts
the resulting MC energy spectrum after the fit to the data points. Only SS events are considered in this example. The data and the smeared MC spectra are each normalized to one. The MC spectrum without energy smearing has an arbitrary normalization. The inset shows the ratio of calibrated data to the smeared MC with the linear fit superimposed (black line).}
 \label{fig:mcfit}
\end{figure}

The calibrated energy \(E\) is obtained as \(E = p_0 + p_1E_r + p_2E^2_r\), where \(E_r\) is the measured energy and \(p_0\), \(p_1\) and \(p_2\) are empirical constants.
The measured energy is assumed to follow a conditional
Gaussian distribution, with the following energy-dependent resolution: \(\sigma^{2}(E)=\sigma_{\text{elec}}^{2}+bE+cE^{2}\), where \(\sigma_{\text{elec}}\) is interpreted as the electronic noise contribution, \(bE\) represents statistical fluctuations in the ionization and scintillation, and \(cE^2\) is assumed to be a position- and time-dependent broadening.
In this analysis, both the energy scale and resolution are determined by fitting the full shape of true energy spectra, as generated by MC, to the corresponding calibration data. This minimizes potential biases caused by determining peak positions and widths using simplified analytical fit models. It allows one to constrain the calibration parameters by utilizing all mono-energetic gamma lines simultaneously in the presence of complex backgrounds due to Compton scatters, summation peaks, and passive detector materials. Before the fit, the MC energy spectrum does not include effects of the energy smearing observed with the detector (Figure~\ref{fig:mcfit}). In the fitting process, the simulated energy spectra from MC are folded with the measured detector response. The resolution and calibration parameters are fitted simultaneously using a maximum likelihood fit. Similar procedures were used in our previous analyses (\cite{Albert:2014,Albert:2014b}) to calculate only the resolution parameters. The available source calibration data allows the above fit to be performed on a weekly basis under the assumption of \(c\) = 0 and \(p_2\) = 0. However, comprehensive calibration data acquired less frequently, but with increased statistics, is used to provide a time-averaged quadratic correction to the weekly calibration
parameters. This correction is measured at the sub-percent level. The correction, as well as the time-averaged resolution parameters used in this analysis, is determined by maximizing a likelihood function that takes into account the livetime of physics runs.

Probability density functions (PDFs) for signal and background components are created using a Monte Carlo simulation. 
 Compared to the previous analyses, the MC was improved by substituting simplified modeling of the noise in the signal waveforms with real noise traces sampled from the data and by adjusting the amplitude of simulated signals to better match the data. This resulted in improved agreement between data and MC of the energy threshold for full position reconstruction and improved agreement in average SS fraction. A \(\sim\)5\% discrepancy in the shapes of the energy distributions, however, remained. This discrepancy, which is included as a systematic error, manifests itself as an excess of SS events in the data over MC at energies around 1 MeV that gradually and linearly decreases with energy, eventually turning into a deficit (Figure~\ref{fig:mcfit}).
The PDFs are functions of the two observables: energy and standoff distance (SD). SD is defined as the distance between a charge deposit and the closest material that is not liquid xenon, other than the cathode, emphasizing separation between events originating outside and inside of the chamber. For a multi-site event, the smallest standoff distance among multiple charge clusters is used to define SD for the event.
Components comprising the overall PDF model are the same as in~\cite{Albert:2014b} with the neutrino-mediated \(0\nu\beta\beta\) signal replaced by a Majoron-emitting decay. The parameters of the overall model are the event counts and SS fractions of individual components, and three variables representing normalization terms. The first normalization term is common to all components and is subject to uncertainty due to event reconstruction and selection efficiencies. The second normalization term is specific to the Majoron-emitting decay component and incorporates uncertainty due to discrepancy in shapes of Monte Carlo and data distributions. The third normalization term incorporates uncertainty due to background model incompleteness and applies to background components in the fit. The normalization terms are included in the PDF in a way analogous to the one described in~\cite{Albert:2014}.

An important parameter of the PDFs for \(\beta\)-like components (e.g. \(0\nu\beta\beta\chi_0(\chi_0)\)) is the ``\(\beta\)-scale'', which describes possible difference in energy scales of \(\beta\)-like and \(\gamma\)-like (e.g. external backgrounds) events. 
The \(\beta\)-scale variable is defined as an energy independent ratio of \(\gamma\) over \(\beta\) energy scales.  
The \(\beta\)-scale is of particular importance for this analysis because adding a \(\beta\)-like component with continuous energy spectrum, such as \(0\nu\beta\beta\chi_0(\chi_0)\), introduces correlation with the \(2\nu\beta\beta\) component and reduces the accuracy with which both the \(\beta\)-scale and the Majoron components can be determined. While the central values of the \(\beta\)-scale found for each mode, as well as for the case of no Majoron mode, are consistent with 1, the corresponding uncertainty increases the final error on each Majoron-emitting decay rate.

A negative log-likelihood function is formed between the data and the overall PDF with the addition of several Gaussian constraints~\cite{Albert:2014} that incorporate systematic uncertainties determined by stand-alone studies. The following parameters are constrained by their corresponding errors, indicated in parentheses: SS fractions (4\%), activity of radon in the liquid xenon (10\%), common normalization term (8.6\%), Majoron-specific normalization term (16\% for spectral index n=1, 30\% for other Majoron modes), background normalization term (20\%) and relative fractions of neutron-capture related PDF components (20\%). The methodology for determining the systematic errors follows the one described in~\cite{Albert:2014b}. The fit is performed simultaneously for SS and MS events.

\section{Results and conclusion}\label{sec:Results} 
\begin{figure*}[htbp]
 \centering
\includegraphics[width=0.99\textwidth]{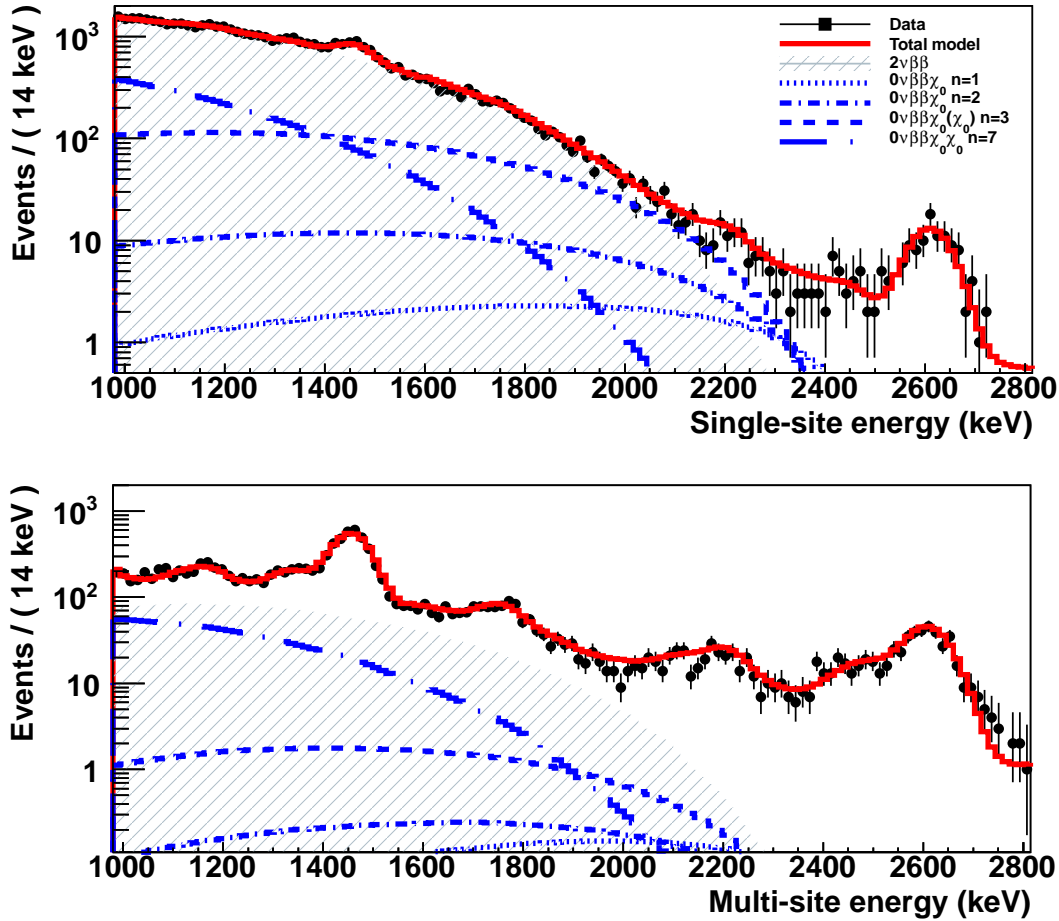}
  \caption{(Color online) SS (top) and MS (bottom) datasets and the best-fit models for the case of the n=1 Majoron fit. SS energy is predominantly populated by \(\beta\)-like events. The most abundant fit component - the \(2\nu\beta\beta\) decay - is shown in hatched gray. The upper 90\% C.L. limits on the number of decays for each of the four Majoron emitting modes are plotted on the figure all at once, as an illustration.}
 \label{fig:fit}
\end{figure*}

A profile likelihood scan is performed for each Majoron-emitting \(0\nu\beta\beta\) decay mode separately. The results are consistent with zero amplitude at less than 1 sigma for Majoron emitting modes with spectral indices 1, 2 and 3, and at \(\sim\)2.2 sigma for n=7, as determined with a toy MC study. As a consistency check, we compare the half-life of the \(2\nu\beta\beta\) decay extracted from the fits with additional Majoron components (added one at a time) to the result published previously~\cite{Albert:2014}. The \(2\nu\beta\beta\) half-life values are consistent within 2-3\% for the Majoron-emitting decay modes with spectral indices 1,2 and 3, and within 12\% for spectral index 7. Given that the uncertainty on the \(2\nu\beta\beta\) half-life in this measurement reaches \(\sim\)8\% due to larger fiducial volume and additional correlation with the \(0\nu\beta\beta\chi_{0}(\chi_{0})\) component, we consider these results to be in good agreement. The robustness of the Majoron fits was also checked against the existence of hypothetical backgrounds not included in the background model, in particular \(^{110m}\)Ag and \(^{88}\)Y, which have gamma lines with energies close to the maxima of some of the Majoron modes. Additional fits were performed for each Majoron mode with each background included in the overall model (one at a time). The contribution of these components was found to be effectively constrained by the multi-site energy distribution, resulting in much less than 1 sigma impact on the Majoron fits. 
Figure~\ref{fig:fit} shows the dataset and the best-fit model for the case of the n=1 Majoron fit. The upper 90\% C.L. limits on the number of decays for each of the four Majoron emitting modes are plotted on the figure all at once, as an illustration. 

Table~\ref{tab:results} summarizes the experimental 90\% C.L. lower limits on half-lives and upper limits on the effective Majoron-neutrino coupling constants. Equation~\ref{eq:coupling} is used to translate the half-lives into coupling constants, where the phase space factors are taken from Table~\ref{tab:psfs}, while the matrix elements are taken from~\cite{Menendez:2009,Simkovic:2009} for the Majoron-emitting decay with n=1, and from~\cite{Hirsh:1996} for other modes. Note that the phase space factor for the n=1 Majoron is a factor of two larger in~\cite{Hirsh:1996} than in~\cite{Doi:1985} and~\cite{Suhonen:1998}. 
The factor of two is the correct choice, as was acknowledged in~\cite{Doi:1988} and is included in Table~\ref{tab:psfs}~\cite{vogel_private:2014}. 

\begin{table*}[t]
\centering
\begin{tabular}{l@{\hskip 0.3in} c@{\hskip 0.3 in} c@{\hskip 0.3 in} c@{\hskip 0.3 in} c@{\hskip 0.3 in}}
\hline\hline
 Decay mode	& Spectral index, n	& Model types & T\(_{1/2}\), yr	& $|\langle g^M_{ee} \rangle|$ \\
\hline
\(0\nu\beta\beta\chi_{0}\) & 1 &	IB, IC, IIB	& \textbf{\(>\)1.2\(\cdot\)10\(^{24}\)}	&	\textbf{\(<\)(0.8-1.7)\(\cdot\)10\(^{-5}\)}	\\
\(0\nu\beta\beta\chi_{0}\) & 2 &	``Bulk''		& \textbf{\(>\)2.5\(\cdot\)10\(^{23}\)}	&	\textbf{--} 					\\
\(0\nu\beta\beta\chi_{0}\chi_{0}\) & 3 &	ID, IE, IID 	& \textbf{\(>\)2.7\(\cdot\)10\(^{22}\)}	&	\textbf{\(<\)(0.6-5.5)}	\\
\(0\nu\beta\beta\chi_{0}\)  & 3 &	IIC, IIF	& \textbf{\(>\)2.7\(\cdot\)10\(^{22}\)}	&	\textbf{\(<\)0.06}				\\
\(0\nu\beta\beta\chi_{0}\chi_{0}\) & 7 &	IIE 		& \textbf{\(>\)6.1\(\cdot\)10\(^{21}\)}	&	\textbf{\(<\)(0.5-4.7)}		\\
\hline\hline
\end{tabular}
  \caption{90\% C.L. limits on half-lives and coupling constants for different Majoron decay models. Spread in coupling constants is due to uncertainty in matrix elements (taken from~\cite{Menendez:2009,Simkovic:2009} for n=1 and from~\cite{Hirsh:1996} for other modes). Phase space factors taken from Table~\ref{tab:psfs}.} 
\label{tab:results}
\end{table*}

The spread in the limits on the coupling constants in Table~\ref{tab:results} for given Majoron mode stems from ambiguity in the matrix elements. The best limits on the coupling constant for the n=1 Majoron from a laboratory experiment come from NEMO-3 (\(\langle g^M_{ee}\rangle ~<(1.6-4.2)\cdot10^{-5}\))~\cite{Barabash:2014} and KamLAND-Zen (\(\langle g^M_{ee}\rangle ~<(0.8-1.6)\cdot10^{-5}\))~\cite{KamLANDZen:2012zzg}. Note that the phase-space integral for the n=1 Majoron used by KamLAND-Zen is about a factor of two smaller than the most up to date value that we used. Therefore, in spite of having a weaker limit on the half-life for the n=1 Majoron (T\(_{1/2}>1.2\cdot10^{24}\) yr at 90\% C.L.), we report a similar limit on the coupling constant ( \(\langle g^M_{ee}\rangle ~<(0.8-1.7)\cdot10^{-5}\)). We note that applying the same phase space factor to the KamLAND-Zen's half-life limit would translate it into the limit on the coupling constant of \(\langle g^M_{ee}\rangle ~<(0.6-1.2)\cdot10^{-5}\). 

In conclusion, we report results from a search for Majoron-emitting double-beta decay modes of \(^{136}\)Xe with two years of EXO-200 data. No statistically significant evidence for this process is found. We obtain limits on the effective coupling constants comparable to the current strongest results by KamLAND-Zen~\cite{KamLANDZen:2012zzg} and  NEMO-3~\cite{Barabash:2014}. The sensitivity to this and other exotic searches with EXO-200 could be improved in the future with a more precise calibration of the possible difference in \(\beta\) and \(\gamma\) energy scales and the reduction of systematic differences between the spectral shapes in data and MC.

\begin{acknowledgments}
    EXO-200 is supported by DOE and NSF in the United States, NSERC in Canada,
SNF in Switzerland, NRF in Korea, RFBR-14-22-03028 in Russia and DFG Cluster of Excellence
``Universe'' in Germany. EXO-200 data analysis and simulation uses resources of
the National Energy Research Scientific Computing Center (NERSC), which is
supported by the Office of Science of the U.S. Department of Energy under
Contract No.~DE-AC02-05CH11231. The collaboration gratefully acknowledges the
WIPP for their hospitality.

\end{acknowledgments}

\bibliography{exo_majoron_analysis}

\end{document}